
\magnification=1200
\documentstyle{amsppt}
\document
\NoBlackBoxes
\centerline{\bf Quantum Knizhnik-Zamolodchikov equations and
holomorphic vector bundles\footnote{\rm To appear in Duke Math.
Journal, June 1993}}

\vskip .20in
\centerline{\bf Pavel I. Etingof}
\vskip .15in
\centerline{Yale University}
\centerline{Department of Mathematics}
\centerline{2155 Yale Station}
\centerline{New Haven, CT 06520 USA}
\vskip .10in
\centerline{e-mail etingof\@ pascal.math.yale.edu}
\vskip .10in
\centerline{July 1992}
\vskip .10in
\baselineskip 10pt

\centerline{\bf Introduction}

In 1984 Knizhnik and Zamolodchikov \cite{KZ} studied the matrix elements of
intertwining operators between certain representations of affine Lie
algebras and found that they satisfy a holonomic system
of differential equations which are now called the Knizhnik-Zamolodchikov
(KZ) equations. It turned out that the KZ equations (and hence,
representation theory of affine Lie algebras) are a source of
a rich variety of special functions. The Gauss hypergeometric function and
its various generalizations were obtained as solutions of the KZ system.

In the recent paper \cite{FR}, Frenkel and Reshetikhin considered
intertwining operators between
representations of quantum affine algebras. It was
shown that their matrix elements satisfy a system of holonomic
difference equations -- a natural quantum analogue of the KZ system.
It was also shown that the solutions of this difference system are very
nontrivial special functions that generalize basic hypergeometric series.
In particular, one of the simplest solutions is expressed in terms of
the $q$-hypergeometric function which was introduced by Heine in the
19th century. This is consistent with the general idea that all
reasonable special functions should come from representation theory,
as matrix elements of certain representations.

The classical KZ system can be regarded as
a local system -- a flat structure in the
trivial vector bundle on the space of sets of
$N$ distinct points on the complex plane. This fact enabled Schechtman
and Varchenko to apply
geometric methods to the study of the KZ equations and obtain a complete
solution for a general simple Lie algebra\cite {SV}. This interpretation can be
extended
to the quantum case \cite{M;R} with the help of
a specially designed discrete analogue of the notion of a
local system. Other than that, the geometric meaning of
the quantum KZ equations has so far remained unclear.

The goal of this paper is to introduce a new geometric interpretation
of the quantum KZ equations. In Chapters 1 and 2, they are linked to certain
holomorphic vector bundles on a product of $N$ equivalent elliptic curves,
naturally constructed by a gluing procedure
from a system of trigonometric quantum $R$-matrices.
Meromorphic solutions of the quantum KZ equations are
interpreted as sections of such a bundle. This interpretation is an
analogue of the interpretation of solutions of the classical KZ equations as
sections of a flat vector bundle. It yields a non-technical proof of
consistency of the quantum
KZ system.

In Chapter 3 it is shown that the
matrix elements of intertwiners between representations of the quantum
affine algebra $U_q(\widehat{\frak sl_2})$ correspond to regular
(holomorphic) sections.

In Chapter 4, Birkhoff's connection matrices for the quantum KZ
equations are interpreted as transition
matrices from one fundamental system of holomorphic sections to another.
They can be used to give an alternative construction of the holomorphic
vector bundle corresponding to the quantum KZ system.

In Chapter 5, the topological structure of the vector bundle
associated to the quantum KZ equations is studied in the special
case of the quantum affine algebra $U_q(\widehat{\frak sl_2})$. The
Chern classes of this bundle are evaluated, and it is shown that they uniquely
determine its topology. The
main tool of this evaluation is the crystal limit $q\to 0$, and its
result shows that the topology of the quantum KZ equations encodes
the structure of the crystal bases in representations of the
quantum group $U_q({\frak sl_2})$.

Chapter 6 treats the special case $N=2$. In this case, one essentially
deals with a vector bundle over an elliptic curve. This bundle is
shown to be semistable (for the case of $U_q(\widehat{\frak sl_2})$)
if the parameters take generic values. The proof makes use of the
crystal limit $q\to 0$.

In Chapter 7, we give a vector bundle
interpretation of the generalized quantum KZ equations for
arbitrary affine root
systems defined recently by Cherednik \cite{Ch}.

For the sake of brevity, the results regarding quantum groups are
formulated and proved for $U_q({\frak sl_2})$. Mostof them can be suitably
extended to $U_q({\frak g})$ where $\frak g$ is an arbitrary simple
Lie algebra.

\centerline{\bf Acknowledgements}

This paper was inspired by the course on the classical and quantum
Knizhnik-Zamolodchikov equations given by my adviser Igor Frenkel
at Yale in the spring of 1992. It is a pleasure for me to thank
Professor Frenkel for helping me build up a necessary background and
guiding my work on this paper.

I would also like to thank Professors David Kazhdan, Nikolai
Reshetikhin, and Alexandre Varchenko for useful discussions.
\newpage

\heading
\bf 1. Trigonometric $R$-matrices and holomorphic vector bundles.
\endheading

Let $V_1,...,V_N$ be a collection of finite-dimensional vector spaces.
Let $W=V_1\otimes V_2\otimes \dots\otimes V_N$.

Let $R\in\text{End}(V_i\otimes V_j)$, $R=\sum_na_n\otimes b_n$,
$a_n\in\text{End}(V_i),b_n\in\text{End}(V_j)$.
Throughout the paper we will use the same notation $
R$ for the operator $\sum_n1_{U_1}\otimes a_n\otimes
1_{U_2}\otimes b_n\otimes 1_{U_3}$
in $\text{End}(U_1\otimes V_i\otimes U_2\otimes V_j\otimes U_3)$ and
the operator $\sum_n1_{U_1}\otimes b_n\otimes
1_{U_2}\otimes a_n\otimes 1_{U_3}$
in $\text{End}(U_1\otimes V_j\otimes U_2\otimes V_i\otimes U_3)$,
where $U_1,U_2,U_3$ are
arbitrary vector
spaces
and $1_U\in\text{End}(U)$
 is the
identity map.

\proclaim{Definition 1.1}
A system of trigonometric quantum $R$-matrices on $V_1,...,V_N$ is a
collection of rational functions
$\lbrace R_{ij}(z)\rbrace$ with values in $\text{End}(V_i\otimes V_j)$
for $1\le i,j\le N$, $i\ne j$ such that

(i) $R_{ij}(z)$ satisfy the quantum Yang-Baxter equation:
$$
 R_{ij}\left(\frac{z_1}{z_2}\right) R_{ik}\left(\frac{z_1}{z_3}\right)
 R_{jk}\left(\frac{z_2}{z_3}\right)= R_{jk}\left(\frac{z_2}{z_3}\right)
 R_{ik}\left(\frac{z_1}{z_3}\right) R_{ij}\left(\frac{z_1}{z_2}\right)
\tag 1.1
$$
in $\text{End}(W)$, $1\le i,j,k\le N$;

(ii) $R_{ij}(z)$ satisfy the unitarity condition
$$
R_{ij}(z)=R_{ji}(z^{-1})^{-1}\tag 1.2
$$
in $\text{End}(W)$, $1\le i,j\le N$;

(iii)
$R_{ij}(z)$ have no poles on the unit circle.
\endproclaim

\proclaim{Definition 1.2}
A collection of linear transformations $A_i\in GL(W)$, $1\le
i\le N$, is compatible to the system of matrices $\lbrace
R_{ij}(z)\rbrace$ if
$$
\gather
A_i R_{jk}(z)= R_{jk}(z)A_i,i\ne j,k,\\
A_iA_j R_{ij}(z)= R_{ij}(z)A_jA_i,\quad 1\le i,j\le N\tag 1.3
\endgather
$$
in $\text{End(W)}$.

\endproclaim

Let $p\in\Bbb C^*, |p|<1$, and let $\Pi$ be the subgroup of $\Bbb C^*$
generated by $p$. Consider the complex torus $T=\Bbb C^*/\Pi$.
We will realize $T$ as an annulus $\lbrace z\in \Bbb
C^*\mid |p|\le |z|<1\rbrace$ with identified boundaries.
Denote
by $T^N$ the direct product of $N$ copies of $T$.

It turns out that  to any system of
trigonometric quantum $R$-matrices on $V_1,...,V_N$ and any collection
of linear transformations compatible to this system one can
naturally associate a holomorphic vector bundle on $T^N$ with fiber $W$.

Partition $T^N$ into $N!$ chambers
$D_s=\lbrace (z_1,...,z_N)\mid |p|< |z_{s(N)}|<
\dots<|z_{s(1)}|<1\rbrace$, where $s$ is a permutation from the
symmetric group $S_N$. These sets are open and disjoint, and their closures
cover $T^N$. Also, they are permuted by the natural action of $S_N$ on
$T^N$: $s_1(D_{s_2})=D_{s_2s_1}$, where $s(z_1,...,z_N)=
(z_{s(1)},...,z_{s(N)})$.

We say that two chambers $D_{s_1}$ and $D_{s_2}$ are adjacent if
$\partial D_{s_1}\cap \partial D_{s_2}$ is a hypersurface of
real codimension 1 in $T^N$. The following trivial lemma classifies pairs
of adjacent chambers.

\proclaim{Lemma 1.1} Chambers $D_{s_1}$ and $D_{s_2}$ are adjacent if and
only if $s_1^{-1}s_2$ is either a transposition of adjacent elements
$t_j=(j,j+1)$, $1\le j\le N-1$, or a cyclic permutation $c^{\pm 1}$,
$c=(N,N-1,...,2,1)$.
\endproclaim

The entire torus can be obtained by gluing the chambers together along
the boundaries between them.
Therefore, in order to define a vector bundle on $T^N$, it suffices to
proclaim it trivial over each of the chambers and prescribe transition
matrices on the boundaries between adjacent chambers.
These boundaries are
$$\Gamma_{s,j}=\lbrace (z_1,...,z_N)\mid |p|< |z_{s(N)}|<
\dots<|z_{s(j+1)}|=|z_{s(j)}|<\dots<|z_{s(1)}|<1\rbrace$$
and
$$
\Sigma_s=\lbrace (z_1,...,z_N)\mid |p|= |z_{s(N)}|<
\dots<|z_{s(1)}|<1\rbrace.
$$
Note that $\Gamma_{s,j}=\Gamma_{st_j,j}$.

Now let $\lbrace R_{ij}(z)\rbrace$ be any system of rational functions
with values in $\text{End}(V_i\otimes V_j)$ satisfying condition (iii),
and let $\lbrace A_i\rbrace$ be a system of linear
transformations compatible to $\lbrace R_{ij}(z)\rbrace$.

On $\Gamma_{s,j}$, define the
transition matrices from $D_s$ to
$D_{st_j}$ to be
$$
T_{s,j}(z_1,...,z_N)=R_{s(j+1)s(j)}\left(\frac{z_{s(j+1)}}{z_{s(j)}}\right)\tag
1.4
$$
On $\Sigma_s$, set the transition matrices from $D_s$ to $D_{sc}$ to
be
$$
Q_s=A_{s(N)}\tag 1.5
$$

These transition matrices take values in $\text{End}(W)$.

\proclaim{Theorem 1.2}
Transition matrices (4) and (5) define a holomorphic vector bundle
on $T^N$ with fiber $W$ if and only if $\lbrace R_{ij}(z)\rbrace$ is a
system of trigonometric $R$-matrices and $\lbrace A_i\rbrace$
are compatible to $\lbrace R_{ij}(z)\rbrace$.
\endproclaim

\proclaim{Remark 1.1} \rm
This fact can be
considered as a geometric interpretation of trigonometric solutions to
the quantum Yang-Baxter
equation with a complex parameter.
\endproclaim

\demo{Proof} First of all, the regularity of $\lbrace R_{ij}(z)\rbrace$ on the
unit circle
is equivalent to the smoothness of the transition functions
 on the boundaries between chambers.
 Thus, it
remains to show that the consistency conditions on transition matrices
reduce to relations (1.1), (1.2) and (1.3).

The consistency conditions apply at every point $P\in T^N$ whose
arbitrarily small neighborhood intersects more than two chambers.
Let $U$ be a small enough neighborhood of $P$. Consider the graph
$G(P)$ whose vertices are connected components of intersections of
chambers with $U$.\footnote{Note that although the chambers are
connected, their intersections with $U$ may consist of several
components.} Two vertices
are connected with an edge if the corresponding two components are
adjacent inside $U$. Thus every edge is labeled with a transition matrix --
a holomorphic $\text{End}(W)$-valued function in
$U$. Hence, a holomorphic in $U$ transition matrix can be associated to
every oriented path in the graph $G(P)$. This matrix is defined to be the
product of the transition matrices labeling the edges of this path in
the opposite order to that prescribed by the orientation.
 The consistency conditions demand that any two paths with the same
beginning and end be labeled with the same transition matrix
(this is equivalent to the uniqueness of analytic continuation of
sections of the bundle in $U$).

Let $P=(z_1,...,z_N)$. First assume that $|z_j|\ne |p|$ for all $j$.
Let $s$ be a permutation labeling one of the chambers intersecting
with $U$. Then there exists an increasing sequence
$1\le j_1<j_2<...<j_r=N$ such that
$$
|z_{s(1)}|=\dots=|z_{s(j_1)}|>|z_{s(j_1+1)}|=\dots=|z_{s(j_2}|>\dots
>|z_{s(j_{r-1}+1)}|=\dots=|z_{s(j_r)}|.\tag 1.6
$$
Obviously, vertices of $G(P)$ correspond to permutations of the
form $s\sigma$, $\sigma\in S(P)$, where $S(P)=S_{j_1}\times
S_{j_2-j_1}\times\dots\times S_{j_r-j_{r-1}}\subset S_N$.
Denote such a vertex by $v(\sigma)$. It is clear that $v(\sigma_1)$
and $v(\sigma_2)$ are connected with an edge if and only if
$\sigma_1=\sigma_2t_j$, $j\ne j_1,...,j_{r-1}$. It follows that a path
in $G(P)$ is a representation of an element $\sigma\in S(P)$ as a
product of the generators $t_j$. It is known from the theory of the
symmetric group that any two such representations can be identified
with the help of relations
$$
\gather
t_jt_{j+1}t_j=t_{j+1}t_jt_{j+1},\tag 1.7\\
t_j^2=1,\tag 1.8\\
t_it_j=t_jt_i,\quad |i-j|>1.\tag 1.9
\endgather
$$
Applying this result to the
transition matrices, we find that the consistency conditions at $P$
are equivalent  to identities (1.1) and (1.2) for the matrices $\lbrace
R_{ij}(z)\rbrace$. Indeed, it is easy to see that
relation (1.7) reduces to the
quantum Yang-Baxter equation,
relation (1.8) yields the unitarity, and relation (1.9)
trivially follows from the definition of transition matrices.

Now assume that (1.6) holds and $|z_{s(j)}|=|p|$ for $j_{r-1}<j\le j_r$.
 Set
$k=j_r-j_{r-1}$.
 The graph $G(P)$ can then be represented as a union
$G(P)=G_0(P)\cup G_1(P)\cup\dots\cup G_k(P)$, where
$G_l(P)$ corresponds to the points in the vicinity of $U$ with
$|z_{s(j)}|$ close to $|p|$ for $j_{r-1}< j\le j_r-l$ and close to $1$
for $j_r-l<j\le j_r$.
Vertices of $G_l(P)$ are labeled by permutations from
$S(P)$ as follows: $v_l(\sigma)\in G_l(P)$ corresponds to the
connected component of $D_{s\sigma c^{l}}$. Two
vertices $v_l(\sigma_1)$ and $v_l(\sigma_2)$ in $G_l(P)$
are connected with an edge if
and only if $\sigma_1c^{l}=\sigma_2c^{l}t_j$, $j\ne l,j_1+l,...,j_{r-1}+l$.
Also, two vertices $v_l(\sigma_1)\in G_l(P)$ and
$v_m(\sigma_2)\in G_m(P)$, $l\ne m$, are connected if and only if
$|l-m|=1$ and $\sigma_1=\sigma_2$. This completely describes the
structure of the graph $G(P)$.

Let $k\ge 1$. For $l\le k-1$ consider the paths
$\pi_1(\sigma,l): v_l(\sigma)\to v_l(\sigma t_j)\to v_{l+1}(\sigma t_j)$
and
$\pi_2(\sigma,l): v_l(\sigma)\to v_{l+1}(\sigma)\to v_{l+1}(\sigma
t_j)$,
$j\ne j_1,...,j_{r-1},j_r-l,j_r-l-1$. These paths have a common beginning and
ending, and the fact that they give the same transition matrix
is expressed by the relation
$R_{\sigma(j+1)\sigma(j)}A_{\sigma(N-l)}=A_{\sigma(N-l)}R_{\sigma(j+1)\sigma(j)}$
which is the first part of (1.3).

Now let $k\ge 2$, and for $l\le k-2$ consider the paths

$\pi^{1}(\sigma,l):v_l(\sigma)\to v_{l}(\sigma t_{N-l-1})\to
v_{l+1}(\sigma t_{N-l-1})\to
v_{l+2}(\sigma t_{N-l-1})$
and

$\pi^{2}(\sigma,l):v_l(\sigma)\to v_{l+1}(\sigma)\to v_{l+2}(\sigma)\to
v_{l+2}(\sigma t_{N-l-1})$,

with the
same beginning and ending.
The fact that they should give the same transition matrix is expressed
by the relation
$$
R_{\sigma(N-l)\sigma(N-l-1)}A_{\sigma(N-l-1)}A_{\sigma_(N-l)}=A_{\sigma(N-l)}A_{\sigma(N-l-1)}R_{\sigma(N-l)\sigma(N-l-1)}
,$$
which is the second part of (1.3).

An elementary combinatorial argument shows that any two
paths in the graph $G(P)$ with the same beginning and ending can be
identified with each other
by replacing subpaths of the form $\pi_1$ by $\pi_2$, $\pi^1$ by $\pi^2$,
and vice versa, and using relations (1.7)--(1.9) when necessary.
This means that once (1.1)--(1.3) are satisfied, any two such paths
are forced to have the same transition matrix.
Thus, we have shown that the consistency conditions on transition
matrices
are equivalent to relations (1.1),(1.2), and (1.3).$\blacksquare$
\enddemo

{}From now on we assume that the matrices $\lbrace R_{ij}(z)\rbrace$ and
$\lbrace A_i\rbrace$ are fixed
and denote the vector bundle constructed above by $E_N$.

Let $\Delta=(\Delta_1,...,\Delta_N)$ be a set of complex numbers.
Introduce a new holomorphic vector bundle $E_N(\Delta)$ obtained from
$E_N$ by twisting. For $\delta\in \Bbb C$ define a line bundle
$L_{\delta}$ on $T$ whose meromorphic sections are functions
$\psi(z)$ meromorphic
in $\Bbb C^*$
such that $\psi(pz)=p^{\delta}\psi(z)$. Clearly, the isomorphism class of
$L_{\delta}$
is determined by $\delta \text{ mod}(1, 2\pi\sqrt{-1}/\log p)$, or,
eqivalently, by the image of $p^{\delta}\in \Bbb C^*$ in $T$.
Let $\theta_j:T^N\to T$,
$1\le j\le N$ be the projection: $\theta_j(z_1,...,z_N)=z_j$, and let
$\Cal B_j(\Delta_j)=\theta_j^*(L_{\Delta_j})$.
$\Cal B_j(\Delta_j)$ are topologically trivial holomorphic line
bundles on $T^N$. Let $\Cal B(\Delta)=\Cal B_1(\Delta_1)\otimes \Cal
B_2(\Delta_2)\otimes\dots\otimes \Cal B_N(\Delta_N)$.
Now for any bundle $\Cal E$ over $T^N$ define a new bundle $\Cal E(\Delta)$ by
$$
\Cal E(\Delta)=\Cal E\otimes \Cal B(\Delta).\tag 1.10
$$

In the next section we will identify solutions of the quantum
KZ equations with sections of $E_N(\Delta)$.
\vskip .3in

\heading
\bf 2. Quantum Knizhnik-Zamolodchikov equations and meromorphic sections.
\endheading

Let $C=\lbrace (z_1,...,z_N)\in T^N:1>|z_j|>|p|, 1\le j\le N\rbrace$.
Let $\phi(z_1,...,z_N)$ be a meromorphic function in $\Bbb C^{*N}$
with values in $W$. Set
$\psi(z_1,...,z_N)=z_1^{\Delta_1}\dots
z_N^{\Delta_N}\phi(z_1,...,z_N)$.

To the function $\psi$ we can associate a meromorphic section of
$E_N(\Delta)$ over the open set $C$ as follows.

A meromorphic section of $E_N(\Delta)$ over $C$ is a collection of
functions
$\psi_s(z_1,...,z_N)=z_1^{\Delta_1}\dots
z_N^{\Delta_N}\phi_s(z_1,...,z_N)$, $s\in S_N$, where $\phi_s$ is a $W$-valued
meromorphic function on $D_s^c\cap C$ ($D_s^c$ denotes the
closure of the chamber $D_s$), satisfying the consistency
conditions
$$
\phi_{st_j}=T_{s,j}\phi_s\text{ on }\Gamma_{s,j}.\tag 2.1
$$
Let $w$ be the element of
maximal length in $S_N$ ($w(i)=N-i+1$). To construct a section of
$E_N(\Delta)$ from $\psi$, set $\phi_w=\phi$, and
then extend it to the whole set $C$ by applying rule (2.1). That means,
if $s=t_{j_1}\dots t_{j_n}$ and $s_l=t_{j_1}\dots t_{j_l}$, $1\le l\le
n$, then set
$$\phi_{ws}=T_{ws_{n-1},j_n}T_{ws_{n-2},j_{n-1}}\dots
T_{ws_1,j_2}T_{w,j_1}\phi\tag 2.2$$
in $D_{ws}$. It follows from relations (1.1) and (1.2) that the result of
this extension does not depend on the decomposition of $s$ in the
product of $t_j$. Clearly, functions (2.2) satisfy conditions (2.1).
Thus, we have constructed a section of the bundle $E_N(\Delta)$ over $C$.
Denote this section by $\tilde\psi$.

Now let us define the quantum KZ equations -- the main subject of this
paper.

\proclaim{Definition 2.1} The difference equations on a $W$-valued
function $\psi$
$$
\psi(z_1,...,pz_j,...,z_N)=R_{j,j-1}\left(\frac{z_j}{z_{j-1}}p\right)
R_{j,j-2}\left(\frac{z_j}{z_{j-2}}p\right)\dots
R_{j,1}\left(\frac{z_j}{z_1}p\right)\times
$$
$$A_jR_{j,N}\left(\frac{z_j}{z_N}\right)R_{j,N-1}\left(\frac{z_j}{z_{N-1}}\right)\dots
R_{j,j+1}\left(\frac{z_j}{z_{j+1}}\right)\psi(z_1,...,z_j,...,z_N),\quad
1\le j\le N,\tag 2.3
$$
are called the \it quantum Knizhnik-Zamolodchikov equations.\rm
\endproclaim

The following theorem gives a new geometric interpretation of the
quantum KZ equations.

\proclaim{Theorem 2.1}
The section $\tilde\psi$ of the bundle $E_N(\Delta)$ over $C$ extends
to a global meromorphic section of this bundle if and only if the
function $\psi(z_1,...,z_N)$ satisfies the quantum KZ equations.
\endproclaim

\demo{Proof} In order for $\tilde\psi$ to extend, it is necessary and
sufficient that $\psi_s$ defined above satisfy
the additional consistency conditions on the surfaces $\Sigma_s$.
Obviously, it is enough to require consistency only for
$s=wt_jt_{j+1}\dots t_{N-1}$, $1\le j\le N$. In this case, the conditions are
$$
\phi_{sc}(z_1,...,pz_j,...,z_N)=A_j\phi_s(z_1,...,z_j,...,z_N).  \tag 2.4
$$
Using the decomposition $c=t_{N-1} t_{N-2}\dots t_1$ and equations
(2.2) and (1.4),
we obtain from (2.4)
$$
R_{1,j}\left(\frac{z_{1}}{z_jp}\right)
\dots
R_{j-2,j}\left(\frac{z_{j-2}}{z_jp}\right)
R_{j-1,j}\left(\frac{z_{j-1}}{z_jp}\right)
\psi(z_1,...,pz_j,...,z_N)=
$$
$$A_jR_{j,N}\left(\frac{z_j}{z_N}\right)R_{j,N-1}\left(\frac{z_j}{z_{N-1}}\right)\dots
R_{j,j+1}\left(\frac{z_j}{z_{j+1}}\right)\psi(z_1,...,z_j,...,z_N),\tag 2.5
$$
which is equivalent to (2.3).$\blacksquare$
\enddemo

\proclaim{Corollary 2.2} The quantum KZ equations are consistent.
\endproclaim

\demo{Proof} Any holomorphic vector bundle on a compact complex manifold has
nonzero meromorphic sections. Therefore,
equations (2.3) have nonzero solutions, which implies that they are
consistent.$\blacksquare$
\enddemo

\proclaim{Definition 2.2} Let us say that a system of trigonometric
$R$-matrices $\lbrace R_{ij}(z)\rbrace$ is regular if
for every pair $i,j$ such that $i\ne j$

(i)
$R_{ij}(z)$ is regular and nondegenerate at the origin
and infinity;

(ii) $R_{ij}(z)$ is regular outside the unit circle
and $R^{-1}_{ij}(z)$ is regular inside the unit circle.
\endproclaim

{}From now on we assume that $\lbrace R_{ij}(z)\rbrace$
is regular and use the notation
$$
\gather
M_j(z_1,...,z_N)=\\
R_{j,j-1}\left(\frac{z_j}{z_{j-1}}p\right)
\dots
R_{j,1}\left(\frac{z_j}{z_1}p\right)A_jR_{j,N}\left(\frac{z_j}{z_N}\right)\dots
R_{j,j+1}\left(\frac{z_j}{z_{j+1}}\right)\tag 2.6
\endgather
$$

Fix $s\in S_N$.
Let
$$
M_j^s=\lim_{z_{s(i)}/z_{s(i+1)}\to\infty,1\le i\le
N-1}M_j(z_1,...,z_N).\tag 2.7
$$
It follows from the consistensy of the quantum KZ equations that
$$[M_i^s,M_j^s]=0\tag 2.8$$ for any pair $i,j$.

\proclaim{Theorem 2.3}\cite {FR}

(i) There exists a matrix solution of the quantum KZ equations
of the form
$$
L_s(z_1,...,z_N)=z_1^{\frac{\log M_1^s}{\log p}}\dots
z_N^{\frac{\log M_N^s}{\log p}}F_s(z_1,...,z_N),\tag 2.9
$$
such that $F_s$ is an $\text{End}(W)$-valued meromorphic function in
$\Bbb C^{*N}$ regular in the region
$|z_{s(1)}|>|z_{s(2)}|>...>|z_{s(N)}|$
with $  \lim_{z_{s(i)}/z_{s(i+1)}\to\infty,1\le i\le
N-1}F_s=1_W$ (this function will be homogeneous of degree 0).

(ii) Any vector solution $\psi$ of the quantum KZ equations regular in
the region $|z_{s(1)}|>|z_{s(2)}|>...>|z_{s(N)}|$
has the
form $L_su$ where $u\in W$.
\endproclaim

\demo{Idea of proof} The solution $L_s$ is given by the following limit:
$$
\gather
L_s(z_1,...,z_N)=\\
\lim_{k_{s(j_1)}-k_{s(j_2)}\to\infty,j_1>j_2}\prod_{j=1}^N\prod_{i=0}^{k_j-1}M_j^{-1}(z_1p^{k_1},...,z_{j-1}p^{k_{j-1}},z_jp^i,z_{j+1},...,z_N)
 \prod_{j=1}^{N}(M_j^s)^{k_j}.\tag 2.10
\endgather
$$
The existence of this limit follows from the results of \cite{Ao}.
\enddemo

\proclaim{Theorem 2.4} Let $\phi$ and $\psi$ be as above.
If $\lbrace R_{ij}(z)\rbrace$ is regular and
$\phi(z_1,...,z_N)$ is holomorphic in the region $1\ge |z_{s(1)}|\ge
|z_{s(2)}|\ge
\dots\ge |z_{s(N)}|\ge |p|$, $s\in S_N$, then $\tilde\psi$
is a holomorphic section of the bundle $E_N(\Delta)$.
\endproclaim

\demo{Proof} We will assume that $s=id$. For an arbirary permutation
$s$, the proof is similar.

First of all, let us show that $\phi(z_1,...,z_N)$ is
holomorphic in the annulus
$\Cal A=\lbrace (z_1,...,z_N)\in \Bbb C^N:|p|\le |z_1|,...,|z_N|\le 1\rbrace $.
Let $J$ be an integer such that all $R_{ij}(z)$ are holomorphic whenever
$|z|\le |p|^J$.
We will use the notation $\hat z_j=p^{J(j-1)}z_j$. Since $\psi$ satisfies the
quantum KZ equations, we have
$$
\phi(z_1,...,z_N)=\text{const}\prod_{j=2}^N\prod_{i=0}^{(j-1)J-1}
M_j^{-1}(z_1,...,z_{j-1},p^iz_j,\hat z_{j+1},...,\hat z_N)\phi(\hat
z_1,...,\hat z_N) \tag 2.11
$$
Since $\lbrace R_{ij}(z)\rbrace$ is regular, $M_j^{-1}(z_1,...,z_N)$
is holomorphic if $z_jp\le z_i$ for $i<j$ and $z_j\le z_i$ for $i>j$. This
implies
that all the factors $M_j^{-1}$ in (2.11) are holomorphic in $z_1,...,z_N$ in
$\Cal A$.
The function
$\phi(\hat z_1,...,\hat z_N)$ is also holomorphic in the annulus, since
$|\hat z_1|<\dots<|\hat z_N|$ whenever $(z_1,...,z_N)\in\Cal A$.
This shows that $\phi(z_1,...,z_N)$ is holomorphic in the annulus.

Now we are in a position to prove the holomorphicity of the section
$\tilde\psi$.
Since $\phi$ is holomorphic in $\Cal A$, clearly $\tilde\psi$ is holomorphic in
$D_w$.
 Let us analytically continue $\tilde\psi$ into the
region $D_{ws}$, $s\in S_N$, along some path. Let $s=t_{m_1}\dots
t_{m_l}$ be a minimal length decomposition of $s$. Then over $D_{ws}$
$\tilde\psi$ is represented by a function of the form
$$
\psi_{ws}(z_1,...,z_N)=R_{i_lj_l}\left(\frac{z_{i_l}}{z_{j_l}}\right)\dots
R_{i_1j_1}\left(\frac{z_{i_1}}{z_{j_1}}\right)\psi(z_1,...,z_N).\tag 2.12
$$
The key property of this decomposition
is that $|z_{i_m}|$ is always greater than $|z_{j_m}|$ in $D_{ws}$.
It follows from the fact that the decomposition of $s$ we used had minimal
length. This property and the regularity of the $R$-matrices imply that
$\psi_{ws}$ is a holomorphic (multivalued) function in $D_{ws}$ which proves
that $\tilde\psi$ is a holomorphic section of $E_N(\Delta)$ .$\blacksquare$
\enddemo

Let $s\in S_N$ be a permutation. Let $d=\text{dim}W$, and
let $u_1^s,...,u_d^s$ be the
basis of $W$ such that $M_j^su_i=\Delta_j^{(s,i)}u_i$, $1\le j\le N$,
$1\le i\le d$ (we assume the generic situation when such a basis
exists). Let $$\psi^{(s,i)}=L_su_i\tag 2.13$$, and let $\tilde\psi^{(s,i)}$ be
the corresponding holomorphic sections of $E_N(\Delta^{(s,i)})$.

The following proposition is a corollary of formula (2.10).

\proclaim{Proposition 2.5} The sections $\tilde\psi^{(s,i)}(z_1,...,z_N)$ form
a
basis of the fiber $W$ everywhere except points where
$\frac{z_{s(j_1)}}{z_{s(j_2)}}p^n$ is a pole of
$R_{s(j_1)s(j_2)}(z)$ for a suitable $n\in \Bbb Z$ and pair of indices
$j_1,j_2$ such that $j_1>j_2$. At such points,
$\tilde\psi^{(s,i)}(z_1,...,z_N)$ are linearly dependent.
\endproclaim

\proclaim{Remark 2.1} \rm
 We can legitimately talk about linear dependence or
independence of $\tilde\psi^{(s,i)}$ despite they are sections of
different bundles, because the projectivizations of all these bundles
are isomorphic to each other.
\endproclaim

\proclaim{Remark 2.2} \rm
Proposition 2.5 shows that $\tilde\psi^{(s,i)}$ play the role of a fundamental
system of sections: they give a coordinate frame of the fiber at
all points of the base except those lying on a finite set of
hypersurfaces in $T^N$ of complex codimension 1.
\endproclaim

\heading
\bf 3. Matrix elements of intertwining operators for $U_q(\widehat {\frak
sl_2})$
and holomorphic
sections.
\endheading

The quantum affine algebra $U_q(\widehat {\frak sl_2})$ is a
Hopf algebra obtained by a standard $q$-deformation of the
universal enveloping algebra of the Kac-Moody algebra $\widehat {\frak
sl_2}$ \cite{D;J1}. As an associative algebra, it is generated by elements
$e_i,f_i,K_i^{\pm 1}, i=0,1$ which satisfy the following relations:
$$
\gather
e_if_i-f_ie_i=\frac{K_i-K_i^{-1}}{q-q^{-1}},\\
e_if_j-f_je_i=0, i\ne j,\\
K_ie_iK_i^{-1}=q^2e_i,\ K_ie_jK_i^{-1}=q^{-2}e_j,\ i\ne j\\
K_if_iK_i^{-1}=q^{-2}f_i,\ K_if_jK_i^{-1}=q^2f_j,\ i\ne j\\
K_iK_j-K_jK_i=0,\\
e_i^3e_j-\frac{q^3-q^{-3}}{q-q^{-1}}e_i^2e_je_i+
\frac{q^3-q^{-3}}{q-q^{-1}}e_ie_je_i^2-e_je_i^3=0,i\ne j,\\
f_i^3f_j-\frac{q^3-q^{-3}}{q-q^{-1}}f_i^2f_jf_i+
\frac{q^3-q^{-3}}{q-q^{-1}}f_if_jf_i^2-f_jf_i^3=0,i\ne j.\tag 3.1
\endgather
$$
The comultiplication in $U_q(\widehat {\frak sl_2})$ is defined by
$$
\align
\Delta (K_i)=K_i\otimes K_i,\\
\Delta (e_i)=e_i\otimes K_i+1\otimes e_i,\\
\Delta (f_i)=f_i\otimes 1+K_i^{-1}\otimes f_i,\tag 3.2
\endalign
$$
and the antipode acts according to
$$
S(e_i)=-e_iK_i^{-1},\quad S(f_i)=-K_if_i,\quad S(K_i)=K_i^{-1}.
\tag 3.3$$
Here $q$ is a complex number. We will assume that $q$ is not 0 and not
a root of unity.

The algebra $U_q(\widehat {\frak sl_2})$ can be extended by adding
elements $D^{\pm 1}$ which satisfy the relations
$$
DK_i=K_iD, De_0=qe_0D, Df_0=q^{-1}f_0D, De_1=e_1D, Df_1=f_1D\tag 3.4
$$
The algebra obtained by this extension is denoted by
$ U_q(\widetilde {\frak
sl_2})$.

Two kinds of representations are defined for $U_q(\widehat {\frak
sl_2})$:
Verma modules $V_{\lambda, k}$ and
finite dimensional evaluation representations $V_{\mu}(z)$.

The Verma module $V_{\lambda,k}$ is generated by a highest
weight vector $v$ satisfying the relations
$$
e_iv=0,\quad K_1v=q^{\lambda}v, \quad K_0v=q^{k-\lambda}v,\quad
k,\lambda\in \Bbb C.\tag 3.5
$$
This module is free over the subalgebra generated by $f_i$ and
should be regarded as a quantum deformation of the Verma
module over $\widehat{\frak sl_2}$.

The evaluation representation $V_{\mu}(z)$
is defined with the help of the quantum group $U_q({\frak sl_2})$
(see e.g. \cite{CP}).
$U_q({\frak sl_2})$ is a Hopf algebra generated
by elements $e,f, K^{\pm 1}$ satisfying the relations
$$
\gather
ef-fe=\frac{K-K^{-1}}{q-q^{-1}},\\
KeK^{-1}=q^2e,\\
KfK^{-1}=q^{-2}f,\tag 3.6
\endgather
$$
in which the comultiplication and the antipode are given by
$$
\gather
\Delta (K)=K\otimes K,\\
\Delta (e)=e\otimes K+1\otimes e,\\
\Delta (f)=f\otimes 1+ K^{-1}\otimes f,\tag 3.7\\
S(e)=-eK^{-1},\\
S(f)=-Kf,\\
S(K)=K^{-1}. \tag 3.8
\endgather
$$
For $z\in \Bbb C^*$ Jimbo {J2} defined the canonical
algebra homomorphisms
$$
p_z: U_q(\widehat {\frak sl_2})\to U_q({\frak sl_2})
$$
as follows:
$$
\gather
p_z(e_1)=e,\quad p_z(f_1)=f,\quad p_z(K_1^{\pm 1})=K^{\pm 1},\\
p_z(e_0)=zf,\quad p_z(f_0)=z^{-1}e,\quad p_z(K_0^{\pm 1})=K^{\mp
1} .\tag 3.9
\endgather
$$

Let $V_{\mu}$ be the finite dimensional irreducible representation of
$U_q({\frak sl_2})$ with the highest weight $\mu$ -- a nonnegative integer.
Such a representation is unique for any $\mu$. The homomorphisms $p_z$
allow us to define
the action of $U_q(\widehat {\frak sl_2})$ in $V_{\mu}$. The obtained
$\mu+1$-dimensional representation of $U_q(\widehat {\frak sl_2})$ is
called the evaluation representation and is denoted by $V_{\mu}(z)$.

The Verma module $V_{\lambda,k}$ can be made a $U_q(\widetilde {\frak
sl_2})$-module by setting $Dv=v$, where $v\in V_{\lambda,k}$ is the
highest weight vector. This condition uniquely determines the action
of $D^{\pm 1}$ in $V_{\lambda,k}$.
Let us say that a vector $w\in V_{\lambda,k}$
is at level $n$ if $Dw=q^nw$. Denote the space of all such vectors by
$V_{\lambda,k}[n]$.
The subspace of top level vectors $V_{\lambda,k}[0]$ is a
$U_q({\frak sl_2})$-subrepresentation isomorphic to the Verma module
$\Cal M_{\lambda}$ with highest weight $\lambda$ over $U_q({\frak sl_2})$.

We are interested in formal expressions of the form
$$
\Phi(z)=\sum_{n\in \Bbb Z}\Phi[n]z^{-n},\tag 3.10
$$
where $\Phi[m]:V_{\lambda,k}[n]\to V_{\nu,k}[n+m]\otimes V_{\mu}(z)$
are linear maps
such that
$$
\Phi(z) au=\Delta(a)\Phi(z) u,\quad a\in U_q(\widehat {\frak sl_2}), u\in
V_{\lambda,k}.\tag 3.11
$$

Affording a slight abuse of terminology, we can say that $\Phi(z)$ is a
$U_q(\widehat {\frak sl_2})$
intertwining operator $V_{\lambda,k}\to V_{\nu,k}\otimes V_{\mu}(z)$.
\footnote{This statement is not quite precise because if $u\in V_{\lambda,k}$
and $z\in \Bbb C^*$ then $\Phi(z)u$ lies in $\hat V_{\lambda,k}\otimes
V_{\mu}(z)$, where $\hat V_{\lambda,k}$ is the completion of
$V_{\lambda,k}$ which allows infinite sums of homogeneous vectors of infinitely
decreasing degree.}
It is easy to show that any such
operator is uniquely determined by its action on top level vectors.
Moreover, it
is clear that if $w\in V_{\lambda,k}$ is top level then it is
enough to know the top level component of $\Phi(z) w$ in order to
retrieve $\Phi(z) w $. This shows that $\Phi(z)$ is uniquely
determined by the map $\Phi[0]:V_{\lambda,k}[0]\to V_{\nu,k}[0]\otimes
V_{\mu}(z)$. This map must be an intertwining operator $\Cal M_{\lambda}\to
\Cal M_{\nu}\otimes V_{\mu}$ over $U_q({\frak sl_2})$. Moreover, for generic
values of $q,k$ this condition is necessary and sufficient for
$\Phi[0]$ to extend to the entire module $V_{\lambda,k}$ \cite{FR}.
{}From now on we consider this generic situation.

Clebsch-Gordan formula tells us that there is a unique nonzero
intertwiner $\Phi^{\lambda,\nu,\mu}$ of form
(3.10) up to a constant if $\nu+\mu-\lambda$ is even, nonnegative, and
not greater than $2\mu$, and no nonzero
intertwiners of this form otherwise.

Let $u\in V_{\mu}(z)^*$. Define the operator
$\Phi^{\lambda,\nu,\mu}(u,z): V_{\lambda,k}\to \hat V_{\nu,k}$,
$\Phi^{\lambda,\nu,\mu}(u,z)w=u(\Phi^{\lambda,\nu,\mu}(z)w)$.

Despite the range of the operator $\Phi^{\lambda,\nu,\mu}(u,z)$ lies
in the completion of the highest
weight module $V_{\nu,k}$, one can form products of such operators.
If $|z_1|>|z_2|>\dots>|z_N|$ then the product
$$
\Phi^{\lambda_1,\lambda_0,\mu_1}(u_1,z_1)\dots\Phi^{\lambda_N,\lambda_{N-1},\mu_N}(u_N,z_N)\tag
3.12
$$
is a well defined linear map: $V_{\lambda_N,k}\to\hat V_{\lambda_0,k}$.

 Let $v_{\lambda_N,k}$ and $v_{\lambda_0,k}^*$ be the
highest weight vector of $V_{\lambda_N,k}$ and the lowest weight vector
of $V_{\lambda_0,k}^*$, respectively, and let
$\Lambda=(\lambda_0,\lambda_1,...,\lambda_N)$. Form the scalar product
$$
\varphi^{\Lambda}(u_1,...,u_N,z_1,...,z_N)=<v_{\lambda_0,k}^*,\Phi^{\lambda_1,\lambda_0,\mu_1}(u_1,z_1)\dots\Phi^{\lambda_N,\lambda_{N-1},\mu_N}(u_N,z_N)v_{\lambda_N,k}>.\tag
3.13
$$
This scalar product is a matrix element of the intertwiner (3.12). We
will regard it as a holomorphic function in $z_1,...,z_N$ in the
region $|z_1|>|z_2|>\dots>|z_N|$ with values in the finite dimensional space
$V_{\mu_1}\otimes\dots\otimes V_{\mu_N}$ and write it
as $\varphi^{\Lambda}(z_1,...,z_N)$.

 Let $h(\lambda)=\frac{\lambda^2+2\lambda}{2(k+2)}$. Let
$\Delta_i(\Lambda)=h(\lambda_{i-1})-h(\lambda_i)$, $1\le i\le N$.
Define a new (multivalued) function
$$
\Psi^{\Lambda}(z_1,...,z_N)=z_1^{\Delta_1(\Lambda)}\dots
z_N^{\Delta_N(\Lambda)}\varphi^{\Lambda}(z_1,...,z_N).\tag 3.14
$$

It turns out that the function $\Psi^{\Lambda}$ is a product of a scalar
function and a solution of the quantum KZ
equations associated with a certain system of trigonometric
$R$-matrices which is described explicitly as follows.

Let $m,n\ge 0$ be integers.
As a $U_q({\frak sl_2})$-module, $V_m\otimes V_n$ decomposes as
$$
V_m\otimes V_n=\oplus_{r=0}^{\min(m,n)}V_{m+n-2r}.
$$
Let $u_m$ be a highest weight vector of $V_m$, and
let $\omega_{mn}^r$ be the highest weight vectors of the components
$V_{m+n-2r}$ such that $(e\otimes
1)\omega^r_{mn}=\omega_{mn}^{r-1}$ for $r\ge 1$, and
$\omega_{mn}^0=u_m\otimes u_n$.
Let $P^r_{mn}:V_n\otimes V_m\to V_m\otimes V_n$ be the $U_q({\frak
sl_2})$-invariant map such that $P^r_{mn}\omega_{nm}^r=\omega_{mn}^r$
and $P^r_{mn}\omega_{nm}^l=0$ if $r\ne l$.

Define the map
$ R^q_{ij}(z):V_{\mu_i}\otimes V_{\mu_j}\to V_{\mu_i}\otimes V_{\mu_j}$ by
$$
R^q_{ij}(z)=\sum_{r=0}^{\min(\mu_i,\mu_j)}\prod_{l=0}^{r-1}
\frac{1-zq^{\mu_i+\mu_j-2l}}{z-q^
{\mu_i+\mu_j-2l}}P^r_{\mu_i\mu_j}\sigma_{\mu_i\mu_j}, \tag 3.15
$$
where $\sigma_{\mu_i\mu_j}: V_{\mu_i}\otimes V_{\mu_j}\to V_{\mu_j}\otimes
V_{\mu_i}$ is the permutation of factors.

Let $W=V_{\mu_1}\otimes\dots\otimes V_{\mu_N}$.
Fix a complex number $\lambda_0$. Let $v_i\in V_{\mu_i}$, $1\le i\le N$, and
let $Kv_i=q^{m_i}v_i$. Set $m=\sum_{j=1}^Nm_j$. Define the operators
$A_i^{q,\lambda_0}:W\to W$ by
$$
A_i^{q,\lambda_0}
(v_1\otimes v_2\otimes\dots\otimes v_N)=q^{(2\lambda_0+2+m)m_i}v_1\otimes
v_2\otimes\dots\otimes v_N.\tag 3.16
$$

\proclaim{Proposition 3.1}\cite {CP}

(i) $\lbrace R^q_{ij}(z)\rbrace$ is a system of trigonometric
$R$-matrices on $V_{\mu_1},...,V_{\mu_N}$, regular if $|q|<1$.

(ii) $\lbrace A_i^{q,\lambda_0}\rbrace$ are compatible to
$\lbrace R^q_{ij}(z)\rbrace$.

\endproclaim

Thus, the matrices $\lbrace R^q_{ij}(z)\rbrace$ and $\lbrace
A_i^{q,\lambda_0}\rbrace$ define a holomorphic vector bundle with fiber
$W$. We will denote this bundle
by $E_N^{q,\lambda_0}$.

Observe that $E_N^{q,\lambda_0}=\oplus_{r=0}^{\mu}E_{N,r}^{q,\lambda_0}$, where
$E_{N,r}^{q,\lambda_0}$
is the subbundle of $E_N^{q,\lambda_0}$ whose fiber is the subspace
$W_r$ of vectors of
weight $\mu-2r$ in $W$, $\mu=\sum_{j=1}^N\mu_j$.

Because the transition matrices of the bundle $E_{N,r}^{q,\lambda_0}$
depend only on the ratios $z_i/z_j$, this bundle can be obtained from a bundle
over $T^{N-1}$. Indeed, let $\Theta\subset T^N$ be the diagonal:
$\Theta=\lbrace (z,z,...,z)|z\in T\rbrace$, and let $\eta$ be the projection:
 $\eta:T^N\to T^N/\Theta$ (the space $T^N/\Theta$ is
isomorphic to $T^{N-1}$). Denote by $\xi$ the map $T^N\to T$ which
acts according to the formula $\xi(z_1,...,z_N)=z_1z_2...z_N$.
Let $\Delta_0=\frac{(2\lambda_0+2+r)r}{2N(k+2)}$.
Then $E_{N,r}^{q,\lambda_0}=\xi^*(L_{\Delta_0})\otimes \eta^*(\hat
E_{N,r}^{q,\lambda_0})$, where $\hat
E_{N,r}^{q,\lambda_0}$ is a holomorphic vector bundle on $T^N/\Theta=T^{N-1}$.

\proclaim{Proposition 3.2}\cite{FR}  The function $\Psi^{\Lambda}(z_1,...,z_N)$
can be represented in the form
$$
\Psi^{\Lambda}(z_1,...,z_N)=\prod_{i<j}G_{\mu_i\mu_j}(\frac{z_i}{z_j})\psi^{\Lambda}(z_1,...,z_N),\tag
3.17
$$
where $\psi^{\Lambda}$ is a solution of the quantum KZ equations (2.5)
regular in the region $|z_1|\ge|z_2|\ge\dots\ge|z_N|$,
and $G_{\mu_i\mu_j}(z)$ is a scalar-valued meromorphic function in $\Bbb C^*$.
\endproclaim

The functions $G_{\mu_i\mu_j}(z)$ are described in \cite{FR} as certain
infinite products. In this paper, they will not be of further
interest.

Now let $s=t_{j_1}\dots t_{j_n}$ be a permutation. Let
$s_l=t_{j_1}\dots t_{j_l}$, $1\le l\le n$. Define the function
$$
\gather
\varphi^{s,\Lambda}(u_1,...,u_N,z_1,...,z_N)=R^q_{s_{n-1}(j_n)s_{n-1}(j_n+1)}
\dots R^q_{s_1(j_2)s_1(j_2+1)}R^q_{j_1,j_1+1}\times\\
<v_{\lambda_0,k}^*,\Phi^{\lambda_1,\lambda_0,\mu_{s(1)}}(u_{s(1)},z_{s(1)})\dots\Phi^{\lambda_N,\lambda_{N-1},\mu_{s(N)}}(u_{s(N)},z_{s(N)})v_{\lambda_N,k}>.\tag
3.18
\endgather
$$
(Abusing notation, we write $R_{ij}$ instead of $R_{ij}(z_i/z_j)$). Define
$\Psi^{s,\Lambda}$ by
$$
\Psi^{s,\Lambda}(z_1,...,z_N)=z_{s(1)}^{\Delta_1(\Lambda)}\dots
z_{s(N)}^{\Delta_N(\Lambda)}\varphi^{s,\Lambda}(z_1,...,z_N)\tag 3.19
$$
and $\psi^{s,\Lambda}$ by
$$
\Psi^{s,\Lambda}(z_1,...,z_N)=\prod_{s^{-1}(i)<s^{-1}(j)}
G_{\mu_i\mu_j}(\frac{z_i}{z_j})\psi^{s,\Lambda}(z_1,...,z_N),\tag
3.20
$$

\proclaim{Proposition 3.3}\cite{FR}
$\psi^{s,\Lambda}$ satisfies the quantum KZ
equations and is regular in the region
$|z_{s(1)}|\ge|z_{s(2)}|\ge\dots\ge|z_{s(N)}|$.
\endproclaim

Assume $|q|<1$.
Using Proposition 3.1 and the results of Section 2, we deduce

\proclaim{Proposition 3.4}

(i) $\tilde\psi^{s,\Lambda}$ is a holomorphic
section of the bundle $E^{q,\lambda_0}_N(s^{-1}\Delta(\Lambda))$.

(ii) For any $\lambda\in \Bbb C$, there exist exactly $d=\text{dim}W$ vectors
$\Lambda$ such that $\lambda_0=\lambda$ and $\varphi^{\Lambda}\ne 0$. These
vectors can be arranged
in an order $\Lambda_1(s),...,\Lambda_d(s)$ so that
$\tilde\psi^{s,\Lambda_i(s)}=\tilde\psi^{s,i}$, $1\le i\le d$, where
$\tilde\psi^{s,i}$, $s\in S_N$, is defined by (2.13).
\endproclaim

Thus, we have shown that matrix elements of
intertwining operators between representations of the quantum affine
algebra $U_q(\widehat{\frak sl_2})$
can be geometrically interpreted as holomorphic sections of
a certain holomorphic vector bundle. This interpretation remains valid if
$U_q(\widehat{\frak sl_2})$ is replaced with $U_q(\hat{\frak g})$ where
$\frak g$ is an arbitrary simple Lie algebra.

\heading
\bf 4. Connection matrices as clutching transformations.
\endheading

Let $s_1,s_2\in S_N$. Then
the systems of sections $\tilde\psi^{(s_1,i_1)}$ and $\tilde\psi^{(s_2,i_2)}$
are related by a connection matrix $C^{s_1s_2}=\lbrace
c^{s_1s_2}_{i_1i_2}\rbrace$,
where $c^{s_1s_2}_{i_1i_2}$ is a meromorphic section of the bundle
$\Cal B(\Delta^{(s_1,i_1)}-\Delta^{(s_2,i_2)})$:
$$
\tilde\psi^{s_1,i_1}=
\sum_{i_2=1}^dc^{s_1s_2}_{i_1i_2}\otimes\tilde\psi^{s_2,i_2}.\tag 4.1
$$

The connection matrices will have poles since the systems of solutions
$\psi^{(s,i)}$, $1\le i\le d$, are not everywhere linear independent.
According to Proposition 2.5, the poles will be at points where
$\frac{z_{s(j_1)}}{z_{s(j_2)}}p^n$ is a pole of
$R_{s(j_1)s(j_2)}(z)$ for a suitable $n\in \Bbb Z$ and pair of indices
$j_1,j_2$ such that $j_1>j_2$.

Thus, any matrix element of the connection matrix can be written as
a product of powers
of $z_1,...,z_N$ and a rational expression of elliptic functions in
$\log z_1,...,\log z_N$.

Let $s\in S_N$ and $s^{\prime}=s\cdot(ij)$, where $(ij)$ is the
transposition of $i$ and $j$. Let $R^q_{ij}$ and $A_i$ be defined by
(3.15) and (3.16).

\proclaim{Proposition 4.1}\cite{FR}
 There exists a system of meromorphic functions
$B_{ij}(\zeta)$ with values in $Mat_d(\Bbb C)$, $1\le
i,j\le N$ such that $C^{s,s^{\prime}}(z_1,...,z_N)=B_{s(i)s(j)}(\log
\frac{z_{s(i)}}{z_{s(j)}})$. These matrices satisfy the conditions:

(i) the quantum Yang-Baxter equation:
$$
 B_{ij}\left(\zeta_1-\zeta_2\right) B_{ik}\left(\zeta_1-\zeta_3\right)
 B_{jk}\left(\zeta_2-\zeta_3\right)= B_{jk}\left(\zeta_2-\zeta_3\right)
 B_{ik}\left(\zeta_1-\zeta_3\right) B_{ij}\left(\zeta_1-\zeta_2\right);
\tag 4.2
$$

(ii) unitarity:
$$
B_{ij}(\zeta)=B_{ji}^{-1}(-\zeta);\tag 4.3
$$

(iii) double periodicity:
$$
B_{ij}(\zeta+\log p)=B_{ij}(\zeta);\quad
B_{ij}(\zeta+2\pi\sqrt{-1})=LB_{ij}(\zeta)L^{\prime},\tag 4.4
$$
where $L,L^{\prime}\in \text{End}(W)$ are constant diagonal matrices.
\endproclaim

This statement shows that the connection matrices provide
elliptic solutions to the quantum Yang-Baxter equation (elliptic
quantum $R$-matrices).

Now we can give an alternative construction of the bundle $E^{q,\lambda_0}_N$,
using
the connection matrices as clutching transformations.

Let $X$ be a complex analytic space, and let $\lbrace U_i, i\in
I\rbrace$ be an open cover of $X$. Let $\Cal E_i\to U_i$ be holomorphic
vector bundles, and let $\beta_{ij}:\Cal E_i\mid_{U_i\cap U_j}\to\Cal
E_j\mid_{U_i\cap U_j}$ be isomorphisms of holomorphic bundles such
that the consistency conditions $\beta_{ij}\beta_{ji}=\text{id}$ in
$U_i\cap U_j$,
$\beta_{ij}\beta_{jk}\beta_{ki}=\text{id}$ in $U_i\cap U_j\cap U_k$
are satisfied.
Then one can construct a holomorphic vector bundle $\Cal E$ on $X$ by
setting $\Cal E\mid_{U_i}=\Cal E_i\mid_{U_i}$ and
defining the clutching transformation from $U_i$ to $U_j$ to be
$\beta_{ij}$.

Assume that $p^n/q^m\ne 1$ for any nonzero integers $n$ and $m$. Let $s\in
S_N$, and let $H_s$ be the set of all points $(z_1,...,z_N)\in T^N$
such that if $n\in \Bbb Z$ and $j_1>j_2$ then
$\frac{z_{s(j_1)}}{z_{s(j_2)}}p^n$ is not a pole of $R^q_{s(j_1)s(j_2)}(z)$.
The sections $\lbrace\tilde\psi^{s,i},1\le
i\le d\rbrace$ are linearly independent over $H_s$.

\proclaim{Lemma 4.2} $\lbrace H_s,s\in S_N\rbrace$ is an open cover of
$T^N$.
\endproclaim

\demo{Proof} Let  $P=(z_1,...,z_N)\in T^N$. We need to show that
there exists $s\in S_N$ such that $P\in H_s$. Let $Q$ be the infinite
cyclic subgroup in $T$ generated multiplicatively by $q$. Let
$x_1,...,x_r\in T/Q$ be the distinct images of the points
$z_1,...,z_N\in T$ under the homomorphism $h:T\to T/Q$, and let
$X_j=h^{-1}(x_j)$, $1\le j\le r$. Inside $X_j$, the elements are
naturally ordered: for $a,b\in X_j$, we say that $a\preccurlyeq b$ if
and only if $a/b=q^mp^n$, where $m,n\in Z$ and $m>0$. Let $s\in S_N$ be a
permutation such that $z_{s(i)}\preccurlyeq z_{s(j)}$ implies $i\le
j$. Then $P\in H_s$.$\blacksquare$
\enddemo

Now define a holomorphic vector bundle $\Cal E^{q,\lambda_0}_N$ on $T^N$ as
follows.
Set $\Cal E^{q,\lambda_0}_N\mid_{H_s}$ to be isomorphic to
$\oplus_{i=1}^{d}\Cal B(-\Delta^{(s,i)})\mid_{H_s}$, and define the
clutching transformations $$\beta_{s_1s_2}:\oplus_{i=1}^{d}\Cal
B(-\Delta^{(s_1,i)})\mid_{H_{s_1}\cap H_{s_2}}\to \oplus_{i=1}^{d}\Cal
B(-\Delta^{(s_2,i)})\mid_{H_{s_1}\cap H_{s_2}}$$ by
$\beta_{s_1s_2}=C^{s_2s_1}$. It is trivial to check that the
consistency conditions are satisfied.

\proclaim{Proposition 4.3} The bundles $E^{q,\lambda_0}_N$ and $\Cal
E^{q,\lambda_0}_N$ are
isomorphic to each other.
\endproclaim

\demo{Proof}
The system of sections $\lbrace\tilde\psi^{s,i},1\le
i\le d\rbrace$ defines an isomorphism between the bundles $\oplus_{i=1}^{d}\Cal
B(-\Delta^{(s,i)})\mid_{H_s}$
and $E^{q,\lambda_0}_N\mid_{H_s}$. Transition from one system of sections to
another is performed by the connection matrix $C^{s_1s_2}$.$\blacksquare$
\enddemo

\heading
\bf 5. Topology of the quantum Knizhnik-Zamolodchikov equations and
crystal bases.
\endheading

In this section we study the vector bundle $E^{q,\lambda_0}_N$ topologically,
disregarding its holomorphic structure. It turns out that the
topological structure of $E^{q,\lambda_0}_N$ is quite nontrivial and can be
described in terms of the combinatorics of crystal bases in
representations of quantum groups.

Clearly, we may assume, without loss of generality, that
$\text{dim}V_{\mu_j}
\ge
2$ for all $j$. Then $\text{rank}(E^{q,\lambda_0}_N)\ge 2^N$.

Recall some properties of characteristic classes.
Let $\Cal E$ be a vector bundle on $T^N$. Let
$c_k(\Cal E)\in H^{2k}(T^N,\Bbb Z)$, $0\le k\le N$, be the Chern classes of
$\Cal E$ (by convention $c_0=1$). The sum
$c(\Cal E)=\sum_{k=0}^Nc_k(\Cal E)\in \oplus_{k=0}^NH^{2k}(T^N,\Bbb
Z)$  is called the total Chern class of $\Cal E$. It has the property
$c(\Cal E\oplus\Cal F)=c(\Cal
E)c(\Cal F)$ for any two vector bundles $\Cal E$ and $\Cal F$. Also,
if $\Cal E$ and $\Cal F$ are one-dimensional then $c_1(\Cal
E\otimes\Cal F)=c_1(\Cal E)+c_1(\Cal F)$ \cite{Mi}.

Let $f_j(x_1,...,x_N)$ be the elementary symmetric polynomials:
$$
\sum_{j=0}^Nf_jt^{N-j}=\prod_{j=1}^N(t+x_j).
$$
Let $Q_k(y_1,...,y_k)$ be the Newton polynomials defined by
$Q_k(f_1,...,f_k)=\sum_{j=1}^Nx_j^k$.
The Chern character of a vector bundle $\Cal E$ is defined by
$$
\text{Ch}(\Cal E)=r+\sum_{k=1}^N\frac{1}{k!}Q_k(c_1,...,c_k),\tag 5.1
$$
where $r$ is the rank of $\Cal E$. It has the properties
$\text{Ch}(\Cal E\oplus\Cal F)=\text{Ch}(\Cal E)+\text{Ch}(\Cal F)$,
$\text{Ch}(\Cal E\otimes\Cal F)=\text{Ch}(\Cal E)\text{Ch}(\Cal F)$.

\proclaim{Proposition 5.1}
A complex vector bundle over $T^N$ of rank $r\ge N$
is uniquely determined by its Chern classes.
\endproclaim

\demo{Proof}
Two bundles of rank $r\ge N$ over $T^N$ are equivalent if and only if
they are stable equivalent. Therefore, a bundle of rank $\ge N$ is
uniquely determined by its class in the ring $K(T^N)$. A well known theorem of
$K$-theory
(\cite {K, Theorem 5.3.25}), asserts that the Chern character induces
an isomorphism of rings: $K(T^N)\otimes \Bbb Q\to H^{2*}(T^N)\otimes
\Bbb Q$.
The complex $K$-ring of the torus is isomorphic to the even part of its
cohomology ring. This fact follows from the rule of evaluation of the
$K$-ring of a product of two spaces (see Prop 4.3.24 in \cite {K}).
 Therefore, $K(T^N)$ is a torsion free abelian group, which implies
that the natural map $K(T^N)\to  K(T^N)\otimes \Bbb Q$ is a monomorphism.
Thus, the composed map $\text{Ch}:K(T^N)\to H^{2*}(T^N,\Bbb Q)$ is injective,
Q.E.D.$\blacksquare$
\enddemo

Since $\text{rank}(E^{q,\lambda_0}_N)>N$, in order to describe the bundle
$E^{q,\lambda_0}_N$
topologically it is enough to calculate $c(E^{q,\lambda_0}_N)$. This turns out
to be
quite simple.

When $q$ goes to $0$, the operators $P^r_{mn}$ defined in Chapter 3
tend to finite limits -- a remarkable phenomenon known as
crystallization \cite{Ka}. This implies that the $R$-matrices
$R^q_{ij}(z)$ have finite limits as $q\to 0$.

It is clear that the topological structure of the bundle
$E^{q,\lambda_0}_{N,r}$ is stable under deformations and hence
independent of the values of $q$ and
$\lambda_0$ (as long as $|q|<1$). Therefore, $E_N^{q,\lambda_0}$ is
topologically equivalent to the bundle
$\tilde E^q_N=\oplus_{r=0}^{\mu}E_{N,r}^{q,-1-r/2}$.
As $q\to 0$, the transition matrices of
the bundle $\tilde E^q_N$ have finite limits:
$R_{ij}^q\to R^0_{ij}$, $A_i^{q,-1-r/2}\to 1$ on $W_r$. Therefore,
there exists a limiting bundle $\tilde E_N^0$, with the fiber $W$, defined
by the matrices
$R_{ij}(z)=R_{ij}^0(z)$ and $A_i=1$ according to (1.4) and (1.5), and
the bundle $E_N^{q,\lambda_0}$ is topologically equivalent to $\tilde E_N^0$
for any $q$ and $\lambda_0$.

Let us now describe the limiting $R$-matrix $R_{ij}^0$.
Let $u_m^l=f^lu_m\in V_m$, $0\le l\le m$.

\proclaim{Proposition 5.2}\cite {Ka} Let $q=0$. Then
$P^r_{mn}\sigma (u_m^i\otimes u_n^j)=0$ unless $\min(i,n-j)=r$.
If $\min(i,m-j)=r$ then
$$
P^r_{mn}\sigma (u_m^i\otimes u_n^j)=\cases  u^j_m\otimes
u^i_n,&i+j\le m\\ u^{m-i}_m\otimes u^{2i+j-m}_n,&m< i+j\le n\\
u^{n-m+j}_m\otimes u^{m-n+i}_n,&n<i+j\le n+m\endcases\tag 5.2
$$
if $m\le n$, and
$$
P^r_{mn}\sigma (u_m^i\otimes u_n^j)=\cases  u^j_m\otimes
u^i_n,&i+j\le n\\ u^{i+2j-n}_m\otimes u^{n-j}_n,&m< i+j\le n\\
u^{n-m+j}_m\otimes u^{m-n+i}_n,&m<i+j\le n+m\endcases\tag 5.3
$$
if $m>n$.
\endproclaim

\proclaim{Corollary 5.3} The $R$-matrix at $q=0$ has the form
$$
R_{ij}^0(z) (u_{\mu_i}^k\otimes u_{\mu_j}^l)=\cases
z^{-\min(k,\mu_j-l)}u^l_{\mu_i}\otimes
u^k_{\mu_j},&k+l\le \mu_i\\ z^{-\min(k,\mu_j-l)}u^{\mu_i-k}_{\mu_i}\otimes
u^{2k+l-\mu_i}_{\mu_j},&\mu_i< k+l\le \mu_j\\
z^{-\min(k,\mu_j-l)}u^{\mu_j-\mu_i+l}_{\mu_i}\otimes
u^{\mu_i-\mu_j+k}_{\mu_j},&\mu_j<k+l\le
\mu_i+\mu_j\endcases\tag 5.4
$$
if $\mu_i\le\mu_j$, and
$$
R_{ij}^0(z) (u_{\mu_i}^k\otimes u_{\mu_j}^l)=\cases
z^{-\min(k,\mu_j-l)}u^l_{\mu_i}\otimes
u^k_{\mu_j},&k+l\le \mu_j\\ z^{-\min(k,\mu_j-l)}u^{k+2l-\mu_j}_{\mu_i}\otimes
u^{\mu_j-l}_{\mu_j},&\mu_j< k+l\le \mu_i\\
z^{-\min(k,\mu_j-l)}u^{\mu_j-\mu_i+l}_{\mu_i}\otimes
u^{\mu_i-\mu_j+k}_{\mu_j},&\mu_i<k+l\le
\mu_i+\mu_j\endcases\tag 5.5
$$
if $\mu_i>\mu_j$.
\endproclaim

Let $L=(l_1,...,l_N)$, $0\le l_j\le \mu_j$. Let
$U_L=u_{\mu_1}^{l_1}\otimes\dots\otimes u_{\mu_N}^{l_N}\in W$. The
vectors $\lbrace U_L\rbrace$ form a basis of $W$ (the crystal basis).

\proclaim{Corollary 5.4}
In the basis $\lbrace U_L\rbrace$, the operator $M_j(z_1,...,z_N)$ defined by
(2.7) can be represented in
the form of a product
$$
M_j(z_1,...,z_N)=Y_j(z_1,...,z_N)\pi_j\tag 5.6
$$
where $\pi_j$ is a permutation matrix ($\pi_j\in S_d$,
$d=\text{dim}W$), and $Y_j$ is a diagonal matrix whose eigenvalues are
products of integer powers of $z_1,...,z_N$.
\endproclaim

It is clear that the permutations $\pi_j$ commute with each other. Let
$\alpha_j$ be the order of $\pi_j$ in $S_d$. Let $\hat T^N$ be the
torus $\prod_{j=1}^N(\Bbb C^*/\Gamma_j)$, where $\Gamma_j$ is the
infinite cyclic group multiplicatively generated by $p^{\alpha_j}$.
Let $\beta:\hat T^N\to T^N$ be the natural covering. Consider
the holomorphic
vector bundle $\beta^*(\tilde E_N^0)$ on $\hat T^N$.
Corollary 5.4 implies that this bundle breaks up into a direct sum of
line bundles, and each of these line bundles is a tensor product of
line bundles over factors $\Bbb C^*/\Gamma_j$. This allows us to
calculate $c(\beta^*(\tilde E_N^0))$ using the properties of Chern
classes that we discussed above. Applying the map
$(\beta^*)^{-1}:H^*(\hat T^N,\Bbb Q)\to H^*(T^N,\Bbb Q)$ to
$c(\beta^*(\tilde E_N^0))$,
 we obtain the sought-for total Chern class $c(E_N^0)$.

For the sake of brevity, we implement this calculation under the
assumption that all $V_{\mu_j}$ are the same: $\mu_j=m$ for all $j$.
Then formulas (5.4) and (5.5) undergo a major simplification:
$R_{ij}^0(z)=R^0(z)$, $i\ne j$, and
$$
R^0(z) (u_m^k\otimes u_m^l)=z^{-\min(k,m-l)}u_m^l\otimes u_m^k.\tag 5.7
$$

In the
formulas below, indices $i,j$ are allowed to take values between 1 and
$N$, and operations on them are fulfilled modulo $N$. For instance,
$N+1$ is identified with 1, and $0$ is identified with $N$.

Introduce the notation $I(k,l)=\min(k,m-l)$,
$J(L,j)=\sum_{i=1}^{j-1}I(l_i,l_{i+1})$, and
$J(L)=\sum_{i=1}^NI(l_i,l_{i+1})$.

Substituting (5.7) into (2.6), we obtain
$$
\gather
M_j(z_1,...,z_N) u_m^{l_1}\otimes\dots\otimes u_m^{l_{N-1}}\otimes u_m^{l_N}=\\
z_j^{-J(L)+I(l_{j-1},l_j)}\prod_{i\ne j}z_i^{I(l_{i-1},l_i)}p^{-J(L,j)}
u_m^{l_N}\otimes u_m^{l_1}\otimes\dots\otimes u_m^{l_{N-1}}\tag 5.8
\endgather
$$

This shows that for all $j$ $\pi_j$ is the cyclic permutation
$(123...N)$. Therefore, $\alpha_j=N$ for every $j$. Let
$$
\hat M_j(z_1,...,z_N)=M_j(z_1,...,p^{N-1}z_j,...,z_N)\dots
M_j(z_1,...,pz_j,...,z_N)M_j(z_1,...,z_j,...,z_N)\tag 5.9
$$
The matrices $\hat M_j$ are the transition matrices for the bundle
$\beta^*(\tilde E_N^0)$. On the other hand, as we have mentioned,
these matrices are diagonal:
$$
\hat M_j(z_1,...,z_N)U_L=z_j^{-(N-1)J(L)}\left(\prod_{i\ne
j}z_i\right)^{J(L)}p^{K(L)}U_L,\tag 5.10
$$
where $K(L)$ is an integer.

Let $x_i=\frac{1}{\log |p|}\text{Re}\log z_i$, $y_i=\frac{1}{2\pi
\sqrt{-1}}\text{Im}\log z_i$. Let $[\Omega]\in H^2(T^N,\Bbb Z)$ be the
cohomology
class of the differential form $$\Omega=N\sum_{j=1}^Ndx_j\wedge
dy_j-(\sum_{j=1}^Ndx_j)\wedge(\sum_{j=1}^Ndy_j)\tag 5.11$$

Formula (5.10) immediately yields an expression for $c(\beta^*(\tilde E^0_N))$:
$$
c(\beta^*(\tilde E^0_N))=\prod_{L}(1+J(L)[\Omega]),\tag 5.12
$$
where the product is taken over all $L=(l_1,...,l_N)$ with $1\le
l_j\le m$. This, in turn, gives a formula for $c(\tilde E_N^0)$:
$$
c(\tilde E_N^0)=\prod_{L}(1+\frac{J(L)}{N}[\Omega]). \tag 5.13
$$
\proclaim{Remark 5.1} \rm
Formula (5.12) contains fractions, but it always gives an integral cocycle
-- the total Chern class of a vector bundle.
\endproclaim

Thus, we have proved

\proclaim{Proposition 5.5} If $V_{\mu_i}=V_m$ for $1\le i\le N$ then
$$
\gather
c(E_{N,r}^{q,\lambda_0})=\prod_{L:\sum
l_j=r}(1+\frac{J(L)}{N}[\Omega]),\quad 0\le r\le Nm,\tag 5.14\\
c(E_N^{q,\lambda_0})=\prod_{L}(1+\frac{J(L)}{N}[\Omega]),\tag 5.15
\endgather
$$
for all values of $q$ and $\lambda_0$.
\endproclaim

\heading
{\bf 6. Special case $N=2$.}
\endheading

According to Chapter 3, in the special case $N=2$ the bundle
$E_{N,r}^{q,\lambda_0}(\Delta)$ can be represented as a product:
$$
E_{2,r}^{q,\lambda_0}(\Delta)=\xi^*(L_{\Delta_0+\frac{1}{2}(\Delta_1+\Delta_2)})\otimes \eta^*(\hat
E_{2,r}^{q,\lambda_0}(\frac{\Delta_1-\Delta_2}{2})),\tag 6.1
$$
where $\xi,\eta:T^2\to T$ are defined by
$\xi(z_1,z_2)=z_1z_2$, $\eta(z_1,z_2)=z_1/z_2$,
$\Delta_0=\frac{(2\lambda_0+2+r)r}{4(k+2)}$, and $\hat
E_{2,r}^{q,\lambda_0}$ is a bundle over $T$.
Clearly, it is enough to understand
$\hat E_{2,r}^{q,\lambda_0}$ in order to understand
$E_{2,r}^{q,\lambda_0}$.

\proclaim{Proposition 6.1} $$\gather\text{rank}(\hat
E_{2,r}^{q,\lambda_0})=\rho_0=\cases r+1,&r\le \min(\mu_1,\mu_2)\\
\min(\mu_1,\mu_2)+1,&\min(\mu_1,\mu_2)<r\le\max(\mu_1,\mu_2)\\
\mu_1+\mu_2-r+1,&\max(\mu_1,\mu_2)<r\le\mu_1+\mu_2\endcases \\
\text{deg}(\hat E_{2,r}^{q,\lambda_0})=\delta_0=\frac{1}{2}\rho_0(\rho_0-1)\tag
6.2
\endgather
$$
\endproclaim

This proposition follows directly from formulas (4.4) and (4.5).
It completely describes the topological structure of the bundle
$\hat
E_{2,r}^{q,\lambda_0}$.

The bundle $\hat
E_{2,r}^{q,\lambda_0}$  is
a holomorphic vector bundle over an elliptic curve. Such bundles were
completely classified by M.Atiyah in 1957 \cite{A}. M.Atiyah showed
that the set of indecomposable bundles of a fixed
rank and degree over an elliptic curve $T$ can be identified with $T$: if
$\Cal E$ is an indecomposable bundle then any other indecomposable
bundle $\Cal F$ of the same rank and degree can be represented in the
form $\Cal F=\Cal E\otimes L$, where $L$ is a line bundle of degree $0$.

In 1962 D.Mumford introduced the notions of stability and
semistability for vector bundles over curves \cite{Mu;NS}. A vector bundle
$\Cal E$ over a curve $X$ is called stable if for every proper nonzero
subbundle $\Cal F\subset \Cal E$ $\frac{\text{deg}(\Cal
F)}{\text{rank}(\Cal F)}<\frac{\text{deg}(\Cal E)}{\text{rank}(\Cal
E)}$, and semistable if $\frac{\text{deg}(\Cal
F)}{\text{rank}(\Cal F)}\le\frac{\text{deg}(\Cal E)}{\text{rank}(\Cal
E)}$. If $\Cal E(t)$ is a holomorphic family of vector bundles on $X$
parametrized by an analytic variety $\Cal T$ then the sets
$\lbrace t\in \Cal T|E(t)\text{ is stable}\rbrace$ and
$\lbrace t\in \Cal T|E(t)\text{ is semistable}\rbrace$ are Zariski
open in $\Cal T$. However, if $X$ is an elliptic curve, stable bundles
of rank $\rho$ and degree $\delta$ over $X$ exist if and only if
$\rho$ and $\delta$ are relatively prime. On the contrary, semistable
bundles over $X$ exist for any values of $\rho$ and $\delta$. Thus,
a generic vector bundle over an elliptic curve is semistable.

Here are some properties of stable and semistable bundles:

\proclaim{Lemma 6.2}

(i) If $\rho$ and $\delta$ are relatively prime, every
semistable bundle of rank $\rho$ and degree $\delta$ is stable.

(ii) If $\Cal E$ is a stable (semistable) bundle and $L$ is a line bundle
then $\Cal E\otimes L$ is stable (semistable).

(iii) If $L_1,...,L_n$ are line bundles of the same degree then the
bundle $\Cal E=L_1\oplus\dots\oplus L_n$ is semistable.

(iv) Let $\Cal E$ be a vector bundle over a curve $X$,
and let $\gamma:\hat X\to X$ be a finite covering. If
$\gamma^*\Cal E$ is a stable (semistable) vector bundle over $\hat X$
then $\Cal E$ is also stable (semistable).
\endproclaim

\demo{Proof} Properties (i) and (ii) are obvious.

(iii) Let $\text{deg}(L_j)=\delta$. Let $\Cal F$ be a subbundle of
rank $\rho$ in $\Cal E$. Then $\Lambda^{\rho}\Cal F$ is a line bundle
embedded in $\Lambda^{\rho}\Cal E$. Obviously, $\Lambda^{\rho}\Cal E=
\oplus_iB_i$, where $B_i$ are line bundles of degree $\rho\delta$.
This implies that for at least one $j$ the projection
$\Lambda^{\rho}\Cal F\to B_j$ is not zero. This projection
is a nonzero regular section of the bundle $\Lambda^{\rho}\Cal
F^*\otimes B_j$. Thus, $\text{deg}(\Lambda^{\rho}\Cal
F^*\otimes B_j)\ge 0$, which implies that $\text{deg}(\Cal F)\le \rho\delta$.

(iv) Pulling back by $\gamma$ does not change the rank of the
bundle and multiplies its degree by the number of sheets of the covering.
Therefore, if $\Cal F\subset \Cal E$ violates the stability
(semistability) condition for $\Cal E$, so does $\gamma^*\Cal F$ for
$\gamma^*\Cal E$.$\blacksquare$
\enddemo

\proclaim{Proposition 6.3} The bundle $\hat E_{2,r}^{q,\lambda_0}$ is
semistable for generic values of $q$ and $\lambda_0$.
\endproclaim

\proclaim{Remark 6.1} \rm
If $\hat E_{2,r}^{q,\lambda_0}$ is semistable then, according to Lemma
6.2, (ii), so is $\hat
E_{2,r}^{q,\lambda_0}(\Delta)$ for every complex number $\Delta$.
\endproclaim

\demo{Proof} First let us show that for a generic value of $q$ the
bundle $\hat E_{2,r}^{q,-r-1/2}$ is semistable. Since the set of such
values is automatically Zariski open in the unit disk, we only need to
show that there exists at least one such value. We will prove it for
$q=0$.

Let $\hat T=\Bbb C^*/\Pi^2$, where $\Pi^2$ is the
infinite cyclic subgroup multiplicatively generated by $p^2$.
Let $\beta:\hat T\to T$ be the natural 2-sheeted covering.
Formulas (5.4) and (5.5) show that the bundle $\beta^*
\hat E_{2,r}^{0,-r-1/2}$ is a direct sum of $\rho_0$ line bundles of degree
$\rho_0-1$. By Lemma 6.2, (iii), this
bundle is semistable. Then Lemma 6.2, (iv) ensures the semistability
of $\hat E_{2,r}^{0,-r-1/2}$.

We know that the set of pairs $(q,\lambda_0)$ for which
the bundle $\hat E_{2,r}^{q,\lambda_0}$ is semistable is Zariski open
in $\Cal D_0\times \Bbb C$, $\Cal D_0$ being the open unit disk
punctured at $0$. This set is also nonempty because it contains almost every
point of the form $(q,-r-1/2)$. Hence, it contains almost every point
of $\Cal D_0\times \Bbb C$.$\blacksquare$
\enddemo

\proclaim{Corollary 6.4} If $r=1$ then
$\hat E_{2,r}^{q,\lambda_0}$ is stable, and hence indecomposable for
generic values of $q$ and $\lambda_0$.
\endproclaim

\demo{Proof} The bundle $\hat E_{2,1}^{q,\lambda_0}$ has degree 1 and
rank 2, and it is semistable for generic values of parameters. Hence,
by Lemma 6.2, (i), it is (generically) stable and therefore
indecomposable.$\blacksquare$
\enddemo

\proclaim{Proposition 6.5} For generic $q$ and $\lambda_0$, the vector
bundle $\hat E_{2,r}^{q,\lambda_0}$ is a direct sum of
line bundles if $\rho_0$ is odd, and a sum of indecomposable
2-dimensional bundles if $\rho_0$ is even. Thus, $\hat
E_{2,r}^{q,\lambda_0}$ is almost always a direct sum of stable
bundles.
\endproclaim

\demo{Proof} Fix a complex number $a$ and consider the family
of vector bundles $\hat E_{2,r}^{q,-1-\frac{r}{2}+\frac{a}{\log q}}$,
$|q|<1$. This family is analytic in the unit disc, and at $q=0$ the
limiting bundle is prescribed by (4) and (5) with
$R_{12}(z)=R^0_{12}(z)$, $A_i(u^{m_1}_{\mu_1}\otimes u^{m_2}_{\mu_2})=
e^{2am_i}u^{m_1}_{\mu_1}\otimes u^{m_2}_{\mu_2}$. As before,
$\beta^*\lim_{q\to 0} \hat E_{2,r}^{q,-1-\frac{r}{2}+\frac{a}{\log
q}}$ is a direct sum of line bundles. Moreover, it is easy to check
that for a generic $a$, all these line bundles are pairwise
non-isomorphic. This shows that such a decomposition takes place for $\beta^*
\hat E_{2,r}^{q,\lambda_0}$ for generic values of $q$ and $\lambda_0$.

Now assume that $\hat E_{2,r}^{q,\lambda_0}$ is semistable but
not a direct sum of
stable bundles. Let $\hat E_{2,r}^{q,\lambda_0}=I_1\oplus\dots\oplus
I_s$, where $I_k$ are indecomposable. Suppose that $I_k$ is unstable
for some $k$. Then, according to \cite{A}, $I_k=\Cal E\otimes \Cal F$
where $\Cal E$ is a
stable bundle (of rank 1 if $\rho_0$ is odd and of rank 2 if $\rho_0$ is even),
and $\Cal F$ is an indecomposable
 degree zero bundle on $T$ with $\text{rank}(\Cal F)>1$. Therefore, we have
$\beta^*I_k=\beta^*\Cal E\otimes\beta^*\Cal F$. We know that
$\beta^*\Cal E$ is either a line bundle or a direct sum of two line
bundles, whereas $\beta^*\Cal F$ is obviously indecomposable. Thus,
$\beta^*I_k$ is not a direct sum of line bundles. This does not
happen for generic values of the parameters.$\blacksquare$
\enddemo

\proclaim{Remark 6.2} \rm
These statements indicate that the trigonometric $R$-matrices arising from
$U_q(\widehat{\frak sl_2})$, as a rule,
give rise to 'generic' holomorphic bundles over $T$, i.e.
ones corresponding to regular points of the moduli space of bundles.
It would be interesting to check if this property holds for other
simple Lie algebras and for $N>2$.
\endproclaim

\proclaim{Remark 6.3} \rm It seems plausible that propositions 6.3 --
6.5 should hold for {\it arbitrary} values of $q$ and $\lambda_0$, not only
for generic ones. However, a proof of this statement would probably have to
involve a more delicate argument than what is used above.
\endproclaim

\proclaim{Remark 6.4} \rm
It seems to be an interesting problem to separate the matrix
elements of intertwining operators from the whole variety of
holomorphic sections of the bundles
$\hat E_{N,r}^{q,\lambda_0}(\Delta)$ by some
geometric condition. It is not clear how to approach this question
since, for instance, the bundle $\hat E_{2,r}^{q,\lambda_0}(\Delta)$, being
generically semistable, has as many linearly independent regular
sections as its degree, for all values of $\Delta$.
\endproclaim

\heading
{\bf 7. Arbitrary root systems}
\endheading

In the recent paper \cite {Ch} a quantum $R$-matrix $R$ was interpreted
as a 1-cocycle on the semidirect product $\tilde W=W\ltimes P$ of a Weyl group
$W$
of the type $A_n$ and its weight lattice $P$
with coefficients in a certain group $\Cal G$ on which $\tilde W$ operates
by automorphisms. The quantum KZ system can then be viewed as a condition
on an element $g\in\Cal G$ (regarded as a 0-cochain on $\tilde W$ with
coefficients in $\Cal G$) requiring that $dg(b)=R(b)$, $b\in P$, where
$d$ denotes the coboundary operator.

This interpretation is very useful since it can be made a definition
of a quantum $R$-matrix and the KZ system
if $A_n$ is replaced with an arbitrary Dynkin diagram.

Our purpose now is to link the 1-cocycle and the vector bundle
interpretations of the quantum KZ equations together and to extend the
vector bundle interpretation to arbitrary root systems.

Let $W$ be a (finite) Weyl group, and let $P$ be its dual root
lattice. Denote by $\Cal K$ the field of $P$-periodic trigonometric
functions on $P\otimes \Bbb C$ (i.e.rational expressions of
monomials $e^{2\pi \sqrt{-1}\lambda(z)}$, $z\in P\otimes \Bbb
C$, where $\lambda\in Q^{\vee}$, $Q^{\vee}$ being the dual lattice to
$P$).
Let $\Cal G=G(\Cal K)$ where
$G=GL(H)$ and $H$ is a finite-dimensional representation of
$\tilde W=W\ltimes P$.

Let $\tau\in \Bbb C^+$. We define the action of $\tilde W$
on $\Cal G$ by
$(wb\circ g)(z)=wg(w^{-1}z-b\tau)w^{-1}$, $b\in P$, $z\in P\otimes \Bbb
C$, $w\in W$.

Let $S_0\subset P\otimes \Bbb R$ be the set of points
with a nontrivial stabilizer in $\tilde W$. Let $S=P\otimes
\Bbb R\oplus \tau
S_0\subset P\otimes \Bbb C$.
Let $\Cal G_0$ be the subgroup of the
elements of $\Cal G$ that are regular at $S$.
Clearly, the action of $\tilde W$ maps elements of $\Cal G_0$ to
(other) elements of $\Cal G_0$.

\proclaim {Definition 7.1} A trigonometric quantum $R$-matrix on $H$
is a $W$-invariant 1-cocycle $R$ on $\tilde W$ with coefficients in $\Cal G_0$
with
the described action of $\tilde W$: for $x,y\in \tilde W$
$R(xy)=(y^{-1}\circ R(x))R(y)$ in $\Cal G_0$.
\endproclaim

Trigonometric $R$-matrices defined in Chapter 1 (for $V_i=V$, $1\le
i\le N$) are obtained
in the special case $H=V^{\otimes N}$, $W=S_N$, and the action of
$\tilde W$ in $H$ factorizes through the action of $W$ in $H$ by
permutations of factors.

Let $\tilde{\Cal K}$ be the field of meromorphic $P$-periodic
functions in $P\otimes \Bbb C$, and let $\tilde{\Cal G}=G(\tilde{\Cal
K})$. Let $\tilde{\Cal G_0}$ be the subgroup of the elements of
$\tilde{\Cal G}$ that are regular at $S$. The action of $\tilde W$ in
$\tilde{\Cal G_0}$ is the same as in ${\Cal
G_0}$.

\proclaim {Definition 7.2} Let $g\in\tilde{\Cal G_0}$. The equations
$(dg)(b)=R(b)$ with respect to $g$, for all $b\in P$, where $d$
is the coboundary operator, are called the quantum KZ system for $R$.
\endproclaim

Again, the quantum KZ equations considered in Chapter 2 come about
in the special case when $H=V^{\otimes N}$
and $W=S_N$.

Now, given a 1-cocycle $R$, we will construct a $W$-equivariant
holomorphic vector bundle on the torus
$T(W)=P\otimes \Bbb C/(P\oplus\tau P)$.

The set $S$ partitions the space $P\otimes \Bbb C$ into
infinitely many bounded
chambers $D_u$ labeled by elements $u\in \tilde W$, so that $u_2\in
\tilde W$ maps $D_{u_1}$ to $D_{u_1u_2}$. If two chambers $D_{u_1}$
and $D_{u_2}$ are adjacent (i.e the intersection of their boundaries
has codimension 1) then $u_2=u_1s$ where $s\in \tilde W$ is a
reflection. Consider the open cover of $P\otimes \Bbb C$ by the
union of small neighborhoods of the chambers $D_u$.
Let us define the transition matrix in the vicinity of the boundary
between $D_{u_1}$ and $D_{u_2}$ (from $D_{u_1}$ to $D_{u_2}$) to be
$$
T_{u_1,s}(z)=(u_1^{-1}\circ R(u_1su_1^{-1}))(z),
\quad z\in P\otimes \Bbb C, \tag 7.1
$$
where the right hand side is regarded as a $GL(H)$-valued function.
By construction, this matrix is regular near the boundary.
Because $R$ is a 1-cocycle on $\tilde W$,
the system of matrices $T_{u,s}$ is a holomorphic
$GL(H)$-valued Cech 1-cocycle on
$P\otimes\Bbb C$. It is easy to check that this
Cech cocycle is invariant
under the action of $\tilde W\ltimes (P\oplus\tau P)$.
Since every Cech cocycle automatically defines a vector bundle,
we obtain a holomorphic vector bundle with fiber $H$ on
$P\otimes \Bbb C$, which descends to a holomorphic bundle over the torus
$T(W)=P\otimes \Bbb C/(P\oplus\tau P)$.

In this setting, meromorphic  solutions of the quantum KZ system are
interpreted as meromorphic sections of the constructed vector bundle,
since geometrically representation of a 1-cocycle as a coboundary of
something is equivalent to constructing a global fundamental system of
sections of the corresponding bundle.

In particular, in the special case $H=V^{\otimes N}$
and $W=S_N$, this scheme gives the vector bundle $E_N$ constructed in
Chapter 1 (assuming $A_1A_2\dots A_N=1$) pushed forward to the
$N-1$-dimensional torus $T^N/\Theta$, where $\Theta$ is the main
diagonal.
This push-forward is possible since if  $A_1A_2\dots A_N=1$ then
$E_N$ is invariant under diagonal translations.

\end